\documentclass[letterpaper,twocolumn,english,rmp,floatfix]{revtex4}
\usepackage[T1]{fontenc}
\usepackage[latin1]{inputenc}
\usepackage{graphicx}

\makeatletter

\usepackage{babel}
\makeatother

\begin{document}

\title[Laser Current Driver]{An Ultra-High Stability, Low-Noise Laser Current Driver with Digital
Control}

\author{Christopher J. Erickson, Marshall Van Zijll, Greg Doermann, Dallin
S. Durfee}

\affiliation{Department of Physics and Astronomy}

\address{Brigham Young University, Provo, UT 84602}

\email{dallindurfee@byu.edu}


\begin{abstract}
We present a low-noise, high modulation-bandwidth design for a laser
current driver with excellent long term stability. The driver improves
upon the commonly-used Hall-Libbrecht design. The current driver can
be operated remotely by way of a micro-processing unit, which controls
the current set point digitally. This allows precise repeatability
and improved accuracy and stability. It also allows the driver to
be placed near the laser for reduced noise and for lower phase lag
when using the modulation input. We present the theory of operation
for our driver in detail, and give a thorough characterization of
its stability, noise, set point accuracy and repeatability, temperature
dependence, transient response, and modulation bandwidth. 
\end{abstract}
\maketitle

\section{Introduction}

Diode lasers have come to play a fundamental role in experimental
physics. Their low cost, efficiency, and small size, among other things,
have made them a standard part of many experiments \citep{Wieman91ud}.
While many turn-key extended-cavity diode laser (ECDL) systems are
presently available, they are typically much more expensive than the
cost of the parts required to build a custom system, and they are
not easily serviced when problems arise. Also, it is difficult to
modify them for special purposes, and they do not provide the extreme
stability and narrow line-widths needed for many applications. As
such, many researchers continue to design and build custom ECDLs.
In this paper we discuss the design and performance of a precision
current driver, one of the key components in any ECDL system.

The laser current controller we present was designed to meet the specifications
needed for the master laser of an optical frequency standard, one
of the most demanding ECDL applications. Optical frequency standards
require a laser with a line-width of a few kHz or less. To achieve
this, these lasers need to have the current stabilized to within a
few tens of nano-amperes rms over relevant time scales, and typically
must be locked to a high-finesse cavity. To simplify the lock to the
cavity, it is desirable for the laser current driver to have a modulation
input with enough bandwidth to allow feedback through the current
driver. This reduces complexity and reduces the chance of destroying
the diode when compared to applying feedback directly to the diode.

Our current driver has several unique design features. The noise and
stability of our current driver are as good or better than any laser
current driver we are aware of. The current driver also has a modulation
input with a 3 dB roll off above 20 MHz. This is much larger than
any of the commercial drivers we have seen, and it maintains its high
stability and low noise even while the high-bandwidth modulation circuit
is active and connected to the driver's output. The driver is also
relatively small in size and features remote programming, allowing
the driver to be placed near the laser to remove the negative effects
of long cables. This design is inexpensive enough to be used as a
reliable lab standard or in student lab experiments.

Our design builds upon the Hall-Libbrecht current driver \citep{Libbrecht93al},
a precision current controller design which has been used extensively
in atomic physics labs. We made several design improvements and updated
components. The most radical improvement is the use of a precision
digital-to-analog converter (DAC), rather than a potentiometer, to
program the current set point. This eliminates a major source of noise
and instability. This also makes it possible to program the current
set point remotely, allowing the current driver to be placed in close
proximity to the laser diode. This can significantly reduce current
noise at the laser, and will lower phase shifts when using the modulation
input. The use of a DAC also allows much more accurate control of
the current set point, and makes it possible to return to a previous
current set point with extremely high precision. Our driver utilizes
surface mount technology, allowing us to use the most modern components
and reducing inductance and noise pick-up from longer pin leads. 

In addition to explaining our improvements to the Hall-Libbrecht circuit
we present a much more detailed description and a more comprehensive
and careful performance analysis of our design than was given in \citep{Libbrecht93al}.
This discussion illuminates both our improvements as well as subtleties
present in the Hall-Libbrecht driver which were not discussed in \citep{Libbrecht93al}.

\section{Design and Construction}

\begin{figure*}
\begin{centering}
\includegraphics{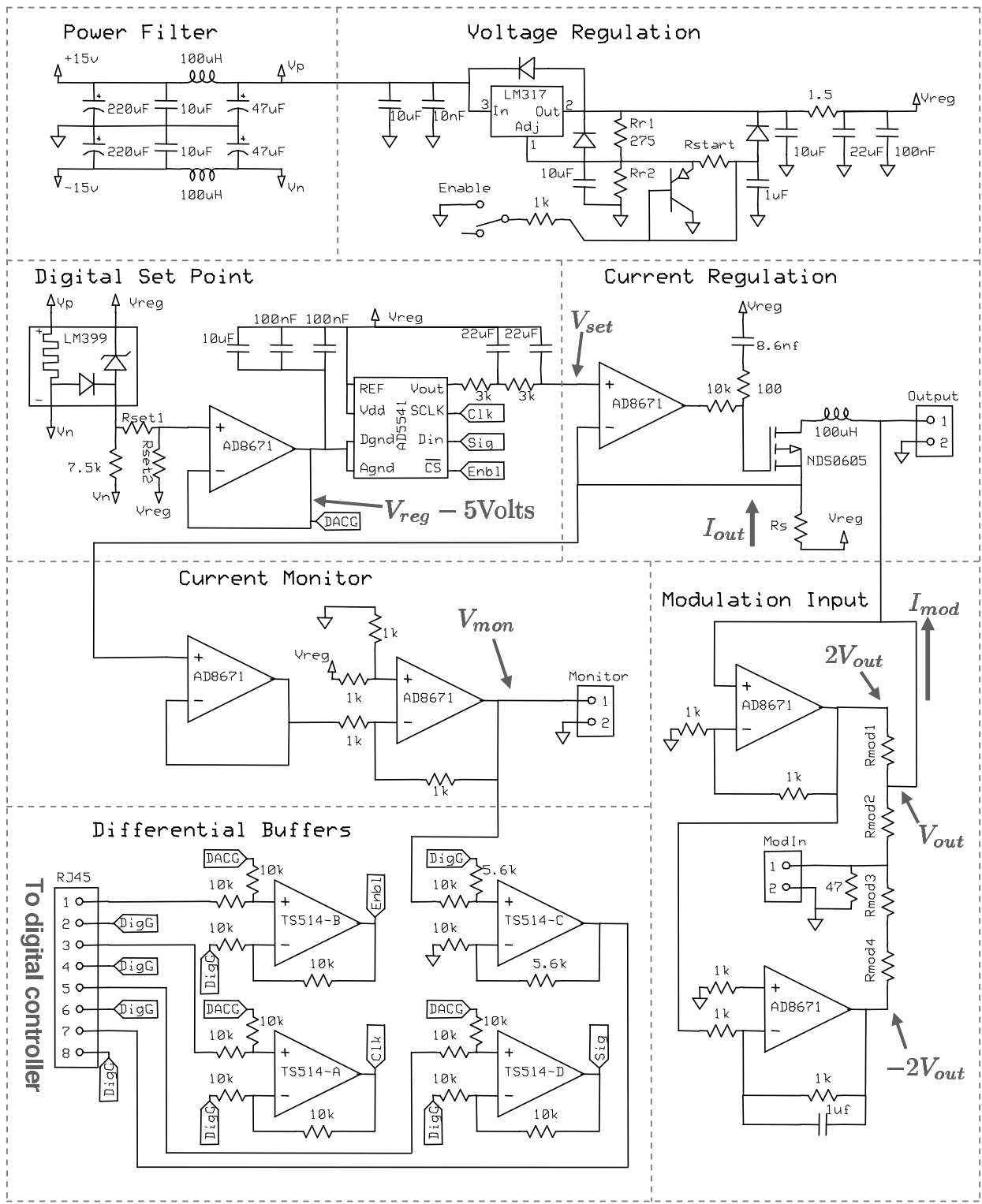} 
\end{centering}

\caption{Schematic of the laser diode current driver circuit. The circuit is
divided into separate functions for clarification and easy referencing.
\label{cap:schematic}}

\end{figure*}
In figure \ref{cap:schematic} we have broken the schematic of the
circuit into several sections, which we will address separately. As
presented, the circuit is configured to drive cathode grounded lasers,
but we have also constructed anode-grounded supplies with a few straightforward
modifications. Although not indicated on the schematic, each op-amp
is powered by the filtered voltages $V_{p}$ and $V_{n}$, with a
pair of capacitors connected between power and ground near each power
pin of each op-amp. Surface-mount ceramic capacitors $0.01\,\mu\mbox{F}$
and $10\,\mu\mbox{F}$ in value are used for the op-amps in the {}``modulation
input'' sub-circuit, the quad op-amp in the {}``differential buffers''
sub-circuit, and the op-amp used for the differential amplifier in
the {}``current monitor'' sub-circuit. For the power lines for the
other three op-amps a $0.1\,\mu\mbox{F}$ ceramic capacitor and a
$100\,\mu\mbox{F}$ electrolytic capacitor are used. The circuit is
laid out on a 6.125 x 4.25 inch printed circuit board (PCB) that is
electrically isolated from the cast aluminum box it is mounted in.
All of the resistors, with the exception of the ones specifically
discussed in the sections below, are standard 1/8 Watt thick-film
chip resistors with a temperature coefficient of $\pm200$ ppm/$^{\circ}\mbox{C}$
in a 1206 surface mount package. All of the non-polar capacitors are
ceramic capacitors in a 1206 surface mount package. Other components
are discussed later.

\subsection{Current Regulation}

\label{sub:CurrentRegulation}The heart of the current driver is the
{}``current regulation'' sub-circuit. This sub-circuit is a fairly
standard current source design \citep{ArtOfElectronicsCurrentSource},
but components have been chosen very carefully for optimal performance.
In this sub-circuit a regulated voltage, labeled $V_{reg}$ in Fig.\
\ref{cap:schematic}, is used to drive current through a precision
sense resistor, labeled $R_{s}$, to the output of the driver. The
amount of current flowing out to the laser diode can be determined
from the voltage drop across $R_{s}$. One side of the sense resistor
is connected to the inverting input of an AD8671 precision op-amp,
and the output of the op-amp is connected to the gate of an NDS0605
mosfet. Negative feedback in this circuit causes the mosfet gate voltage
to change, modifying the current flowing through the mosfet, until
the voltage at the output of the sense resistor is equal to the set
voltage, $V_{set}$, applied to the non-inverting input of the op-amp.
In this way the op-amp forces the current to remain fixed at a value
of \begin{equation}
I=\frac{V_{reg}-V_{set}}{R_{s}}\label{eq:Iout}\end{equation}
 regardless of the load impedance. Due to the mosfet's gate capacitance,
the resistors and capacitor on the line between the op-amp and the
mosfet are needed to reduce the bandwidth of the circuit and prevent
oscillations. As was done in \citep{Libbrecht93al}, an inductor is
placed after the mosfet to limit high-frequency current noise and
to de-couple the current regulation electronics from any intentionally-applied
current modulation.

The sense resistor is a key component which needs to be considered
carefully --- stability of the current driver can be no better than
the stability of this resistor. The value for $R_{s}$ should be chosen
to provide a large enough voltage drop for accurate current regulation
at the target current, but small enough such that the voltage drop
across this resistor does not reduce the compliance voltage of the
driver below what is necessary to power the laser diode. When driving
a red diode laser to a maximum current of 100 mA, we used a pair of
100 Ohm Vishay SMR3D precision resistors in parallel for a total resistance
of 50 Ohms. These parts are surface-mount components for reduced noise
pickup and have a low temperature coefficient of just 2 ppm/$^{\circ}\mbox{C}$.
For higher/lower current applications, smaller/larger resistance values
could be used. In applications that don't require the highest stability,
resistors with higher temperature coefficients may be used.

The op-amp we used was the AD8671 from Analog Devices \citep{AD8671}.
Similar to the LT1028 used in the original Hall design \citep{LT1028},
this op-amp has voltage noise levels comparable to the Johnson noise
of a 50 ohm resistor, the AD8671 having slightly better current noise
and slightly poorer voltage noise than the LT1028. In practice, as
well as in numerical SPICE models, we found that the AD8671 was less
prone to oscillate in this circuit than the LT1028, requiring less
severe bandwidth limiting to make the circuit stable. The AD8671 chip
also has a \emph{much} higher input impedance (both common and differential
mode) than the LT1028 with comparable input capacitance. This property
limits current drift due to thermal or other changes in the leakage
current into the op-amp inputs. In addition, the AD8671 does not require
any sort of trim or compensation to achieve low bias and low overshoot
in our application.

Although the AD8671 has a somewhat smaller gain-bandwidth product
(GBP) than the LT1028 used in the original Hall design (10 MHz vs
50 MHz), the filtering capacitors and resistors needed to stabilize
the circuit effectively remove any additional bandwidth that would
be gained by using the LT1028. Furthermore, the inherent low noise
of the circuit and the use of the de-coupling inductor makes higher
bandwidth in this part of the circuit unnecessary. 

It is worth noting here that we have used the AD8671 op-amps throughout
most of the circuit. The Hall design used LT1028 op-amps only in two
\char`\"{}critical\char`\"{} locations, with different op-amps being
used in the modulation input circuit and the monitoring circuit. Since
both of these circuits connect directly to the current output, it
made sense to us to take advantage of the low noise and high input
impedance of the AD8671. Though more expensive than many lower performance
op-amps, they are still a relatively inexpensive part of this circuit.

For the inductor we choose a $100\mu\mbox{H}$ API Delevan series
S1210 ferrite core shielded surface-mount inductor. The NDS0605 mosfet
was chosen for its low on-state resistance and a relatively large
current carrying capacity in a small form factor. It also has a small
input capacitance, small switching delays, and low gate-body leakage.

\subsection{Digital Set Point}

\label{sub:CurrentSetPoint}The set-point voltage, $V_{set}$, which
is applied to the non-inverting input of the op-amp in the {}``current
regulation'' sub-circuit is generated by a precision digital to analog
converter (DAC) . Using a precision DAC rather than a manual potentiometer
removes a significant source of noise and drift in the circuit. In
addition, it affords much greater accuracy and repeatability. We did
an extensive search to find a DAC which would optimize long-term stability.
The chip we settled on is the 16-bit AD5541 \citep{AD5541}. This
chip comes in several performance grades. All grades have the same
stability, temperature coefficient, and repeatability, but differ
in the absolute accuracy of the output. We chose the {}``J'' grade,
mainly because it was carried by our regular electronics supplier,
but lower cost/lower performance and higher cost/higher performance
grades are also available. For applications where the noise and stability
of the current driver are not a great concern, we have included pads
on the PCB where a traditional potentiometer can be used in place
of the DAC.

The AD5541 chip is specified to have a zero-code temperature coefficient
of 0.05 ppm/$^{\circ}\mbox{C}$ and a gain error temperature coefficient
of 0.1 ppm/$^{\circ}\mbox{C}$, both of which are significantly less
than the temperature coefficient of the sense resistors discussed
in Sec.\ \ref{sub:CurrentRegulation}. The accuracy of the {}``J''
grade chip is guaranteed to be at least as good as $\pm2$ times the
least significant bit (LSB), or 2 parts in 65,535. For our test setup
we used a 50 ohm sense resistor resulting in a maximum current output
of 100 mV. This corresponds to an accuracy of $\pm3\,\mu\mbox{A}$,
similar to its measured accuracy discussed in Sec.\ \ref{sub:AccuracyAndRepeatability}. 

The output of the DAC can be set anywhere from the voltage of its
analog ground pin (labeled $DACG$ in Fig.\ \ref{cap:schematic})
up to the voltage applied to its reference input. As was done with
the analog potentiometer used in \citep{Libbrecht93al}, the ground
pin of the DAC in our circuit is not connected directly to board ground.
There are two reasons why this is necessary. The first is that the
output current of the circuit does not depend on the set point voltage
relative to board ground, but relative to $V_{reg}$. As such, if
the voltage $V_{reg}$ drifts relative to board ground, we want $V_{set}$
to drift with it such that $V_{reg}-V_{set}$ remains constant. The
second is that the DAC requires the voltage on its reference to be
no greater than $5\mbox{ V}$ above the voltage of its analog ground
pin. In order to adjust the output current of the device all they
way down to zero 0, it must be possible to make $V_{set}$ at least
as large as $V_{reg}$. This imposes a lower limit on the voltage
applied to the analog ground pin.

In our circuit $V_{reg}$ is applied directly to the DAC's reference
pin, and a voltage of $V_{reg}-5\mbox{ V}$ (labeled $DACG$ in Fig.\
\ref{cap:schematic}) is applied to the DAC's ground pin, giving the
current driver an output current which can range from 0 to $5\mbox{ Volts}/R_{s}$.
The voltage $DACG$, is generated using an LM399 Zener diode connected
to $V_{reg}$, which creates an extremely stable 6.95 V drop. The
resistors $R_{set1}=1\,\mbox{k}\Omega$ and $R_{set2}=2.74\,\mbox{k}\Omega$
make a voltage divider to lower this 6.95 V difference closer to the
required 5 Volts. We then buffer the voltage with an op-amp. Any drift
in the ratio of $R_{set1}$ and $R_{set2}$ will result in a drift
in the current set point, but there is much less sensitivity to thermal
drift in these resistors than in the sense resistor $R_{s}$ because
only the ratio, and not the absolute resistance matters. Both resistors
are made of the same material and are in close proximity such that
they will have similar thermal drifts. For these two resistors we
use Vishay TNPW0805 precision resistors which have an accuracy tolerance
of 0.1\% and a thermal coefficient of 25 ppm/$^{\circ}\mbox{C}$.
The $7.5\mbox{ k}\Omega$ resistor in this sub-circuit is a Vishay
TNPW04027501DT9 precision, low-temperature coefficient resistor. We
expect, however, that using a standard resistor in this location would
not affect the performance of the current driver.

Similar to what was done in \citep{Libbrecht93al}, a pair of low-pass
filters reduce high frequency noise in $V_{set}$. Because the output
current of the circuit depends not on the difference of $V_{set}$
from board ground, but on the difference between $V_{set}$ and $V_{reg}$,
these filters (as well as the low pass filter in the current regulation
sub-circuit) are tied to $V_{reg}$ rather than ground.

\subsection{Voltage Filtering and Regulation}

\label{sub:VoltageReg}The {}``power filter'' sub-circuit generates
stable, filtered voltages for the rest of the circuit. This sub-circuit
is nearly identical to the one in \citep{Libbrecht93al}. We have
added some additional filter capacitors, changed some component values
to optimize performance, and added two protection diodes between the
three pins of the LM317 voltage regulator, as recommended by the manufacturer
\citep{LM317}. We have also used surface-mount components wherever
possible to reduce noise and stray inductance. 

In this part of the circuit, power lines are first filtered by a series
of capacitors and inductors to provide the positive and negative voltages
$V_{p}$ and $V_{n}$ which are used to power all of the op-amps in
the circuit. The filtered power is then regulated with the LM317 adjustable
voltage regulator to generate the extremely stable voltage labeled
$V_{reg}$. This voltage is used to drive current to the laser diode
in the {}``current regulation'' sub-circuit and to generate the
set-point voltage in the {}``digital set point'' sub-circuit as
described above. 

The transistor below the LM317 is part of an enable/slow start mechanism
discussed in \citep{LM317}. This soft-start feature helps protect
the laser diode from damage, and in the case of a power failure it
also allows the digital controller time to reset the current set point
before $V_{reg}$ turns fully on after power is restored. The regulated
output level of the LM317 is nominally equal to \begin{equation}
V_{reg}=1.25\mbox{ V}(1+R_{r2}/R_{r1})\label{eq:Vreg}\end{equation}
When the \emph{Enable} switch is grounded, the current driver is {}``disabled.''
In this case the resistor $R_{r2}$ is placed in parallel with both
the impedance of the transistor and with the series resistance of
the resistor labeled $R_{start}$ and the 1 k$\Omega$ resistor near
the switch. This effectively reduces the value of $R_{r2}$ in Eq.\
\ref{eq:Vreg}. With a proper choice of resistor values it is possible
to make $V_{reg}$ insufficiently large to forward bias the laser
diode when the switch is in this position. 

When the switch is opened, if not for the 1 $\mu$F capacitor, current
would immediately cease to flow through $R_{start}$. This would cause
the voltage at the gate of the transistor to suddenly equal the voltage
at the emitter of the transistor, shutting off current through the
transistor as well, restoring the voltage regulator to its full output
value as if the soft-start circuitry was not present. The presence
of the 1 $\mu$F capacitor prevents the current flowing through $R_{start}$
from abruptly stopping. Instead, the current continues to flow as
the capacitor charges, and the voltage rises on the gate of the transistor
slowly. This causes the voltage regulator's output to gradually increase
to its final value. Because of the non-linear nature of the transistor,
the start time is considerably larger than just the $RC$ time constant
of the resistor $R_{start}$ and the 1 $\mu$F capacitor. Using a
value of $R_{start}=1\mbox{ k}\Omega$ and a MMBT2907A transistor
in our setup, we measured a start-up time constant of 11.5 ms.

The $220\,\mu\mbox{F}$ capacitors are low-leakage radial aluminum
electrolytic capacitors, and the $47\,\mu\mbox{F}$ capacitors are
surface-mount tantalum capacitors. While developing this circuit the
1.5$\Omega$ resistor in the upper right-hand corner of the schematic
overheated, probably due to increased current draw when our LT1028
op-amps oscillated or saturated. While diagnosing this problem, we
installed a Vishay WSC-1 Series 1 Watt surface-mount wire-wound power
resistor. We have experienced no such difficulties since we began
using AD8671 op-amps, and have since replaced this resistor with a
standard 1/8 Watt metal film resistor. Although we don't expect the
choice of resistor to affect performance, one should note that the
wire-wound resistor was still installed while taking the data presented
in Sec.\ \ref{sec:Results}.

\subsection{Modulation Input}

\label{sub:CurrentModulation}The {}``modulation input'' sub-circuit
is the same one used and described in \citep{Libbrecht93al}. The
key components in this circuit are the upper op-amp and the two resistors
labeled $R_{mod1}$ and $R_{mod2}$ (the lower op-amp is part of a
balancing circuit that we describe later). The non-inverting input
of the upper op-amp is connected directly to the output of the current
driver, which is at a voltage $V_{out}$. The op-amp is wired to generate
a gain of $+2$, such that it generates a voltage of $2V_{out}$ at
its output. The output of the op-amp is connected to the current driver's
output through a $\mbox{1 k}\Omega$ resistor, labeled $R_{mod1}$
in Fig.\ \ref{cap:schematic}. The voltage drop across this resistor
is equal to $2V_{out}-V_{out}=V_{out}$, and so the current flowing
through this resistor is $V_{out}/\mbox{1 k}\Omega$. A second identical
resistor, labeled $R_{mod2}$ in Fig.\ \ref{cap:schematic} connects
the output of the laser driver to the modulation input. If a voltage
$V_{mod}$ is applied to this input, the current flowing through this
resistor is $(V_{out}-V_{mod})/\mbox{1 k}\Omega$. The current which
is added to the output of the laser driver is just the difference
in these two currents, $V_{mod}/1\mbox{ k}\Omega$. This injected
current inevitably results in a change of the output voltage of the
current driver, but the op-amp constantly corrects its output to keep
it equal to $2V_{out}$ as $V_{out}$ changes.

In addition to the modulation current, an additional bias current
will be injected by this sub-circuit if the $R_{mod}$ resistors are
not equally matched or if the gains in the two amplifiers are not
correct. For example, if $R_{mod1}$ and $R_{mod2}$ differ by some
small fraction $\epsilon\equiv(R_{mod2}-R_{mod1})/R_{mod1}$, the
modulation current injected into the laser will be \begin{equation}
I_{mod}=\frac{V_{mod}}{R_{mod2}}+\frac{\epsilon V_{out}}{R_{mod_{1}}(1+\epsilon)}.\label{eq:ImodB}\end{equation}
The bias term (the second term in the above equation) will cause current
measurements based on the voltage drop across $R_{s}$ to be inaccurate.
And if resistances change in time, due to thermal effects for example,
this bias current will change, affecting the stability of the current
driver. As such, these resistors need to be well matched and have
a fairly low temperature coefficient.

All of the resistors in the {}``modulation input'' sub-circuit except
for the $47\,\Omega$ impedance matching resistor on the $ModIn$
input are Vishay TNPW0805 surface-mount thin-film chip resistors with
a resistance tolerance of $\pm0.1\mbox{\%}$ and a temperature coefficient
of 25 ppm/$^{\circ}$C. With this accuracy, the bias term in Eq.\
\ref{eq:ImodB} should be no larger than $\sim2\mu\mbox{A}$ when
driving a typical red laser diode (for which $V_{out}\sim2\mbox{ Volts}$).
Because the resistors $R_{mod1}$ and $R_{mod2}$ have the same composition
and are physically close to each other, such that they should be at
similar temperatures, we wouldn't expect temperature to have a large
effect on the matching of the two resistors. Accordingly, the fact
that these resistors have a much larger temperature coefficient than
the current sense resistor discussed in Sec.\ \ref{sub:CurrentRegulation}
should have no significant effect on the stability of the current
driver. 

The lower op-amp in this sub-circuit forces $V_{mod}$ to 0 V when
nothing is connected to the modulation input (alternatively, this
part of the circuit could be left off, and the modulation input could
be shorted to ground or the {}``modulation input'' sub-circuit completely
disconnected from the current output when not in use). This op-amp
generates a voltage of $-2V_{out}$ which is connected to the modulation
input through two more $1\mbox{ k}\Omega$ resistors, labeled $R_{mod3}$
and $R_{mod4}$. It is easy to see that, if these are perfectly matched
to the other two resistors, the voltage at the modulation input will
float to zero when nothing is connected to it. If they are not exactly
balanced, an additional bias current will be injected into the output
of the current driver when the modulation input is floating. Given
the precision of these resistors, we would expect this bias to be
on the order of $1\,\mu\mbox{A}$ and we wouldn't expect that temperature
drift should be an issue.

Although not discussed in \citep{Libbrecht93al}, we have come to
realize that this design allows modulation speeds greater than the
bandwidth of the op-amps. At very high modulation frequencies one
would expect that the op-amp would not be able to follow the changes
in $V_{out}$ induced by the oscillating modulation current. In this
limit, if we assume that the current driver is connected to a load
with impedance $Z$, that the output of the upper op-amp in this sub-circuit
settles to twice the time-averaged output voltage, and that $R_{mod1}=R_{mod2}$,
Eq.\ \ref{eq:ImodB} becomes \begin{equation}
I_{mod}=\frac{V_{mod}}{R_{mod1}}\left(1+\frac{2Z}{R_{mod1}}\right)^{-1}.\label{eq:HFmodlimit}\end{equation}
If $R_{mod1}\gg Z$, then this equation is nearly the same as the
low frequency limit in Eq.\ \ref{eq:ImodB}. If the op-amp is modeled
as an ideal op-amp with a low-pass filter with a time constant of
$\tau$ on its input, it can easily be shown that $I_{mod}$ changes
monotonically from it's low-frequency value of $V_{mod}/R_{mod1}$
to this high frequency limit as the modulation frequency increases.
Using this model, the phase shift of $I_{mod}$ goes to zero in both
the low- and high-frequency limits, with the largest phase shift,
occurring at a frequency of $f=2\pi(R/\tau^{2}(R+2Z))$, being equal
to \begin{equation}
\phi_{max}=\arctan\left(\sqrt{\frac{Z^{2}}{R(R+2Z)}}\right).\label{eq:ModPhiMax}\end{equation}
Once again, if $R_{mod1}\gg Z$, the maximum phase shift in this simplified
model will be very small.

\subsection{Current Monitor and Differential Buffers}

The {}``current monitor'' sub-circuit generates a voltage proportional
to the output current of the driver so that the output can be monitored.
The first op-amp in this sub-circuit buffers the voltage at the low-voltage
side of the sense resistor $R_{s}$ with a high input impedance op-amp.
This decouples the current monitor output from the current regulation
sub-circuit so that measurements can be made without significantly
affecting the output of the current driver. The second op-amp subtracts
this voltage from $V_{reg}$ to produce a voltage relative to the
board's ground equal to $V_{mon}=I_{out}R_{s}$. 

Because the output of the current monitor sub-circuit is referenced
to board ground, attaching a meter to this output could potentially
introduce ground noise. Op-amp {}``C'' in the {}``differential
buffers'' sub-circuit eliminates this problem, decoupling the meter's
ground from the circuit. Similarly, the other three op-amps in this
sub-circuit decouple the current driver's ground from the grounds
of the three digital input lines which program the DAC (labeled $Clk$,
$Sig$, and $Enbl$ on Fig.\ \ref{cap:schematic}). These three op-amps
are also used to re-reference the digital inputs to the ground pin
of the DAC (which is not board ground).

\subsection{Microprocessor}

The circuit can be controlled by any device capable of generating
a 16-bit serial TTL signal as well as a TTL clock and enable signal
\citep{AD5541}. To collect some of the data presented in this paper,
we used a digital output card on a computer for this purpose. We have
also developed a dedicated digital control circuit based on a PIC18F4550
microprocessor. This is a 40-pin USB programmable microprocessor with
35 input/output pins. It is programmed to read in the signal from
a digital encoder, and subsequently output a value to the DAC. The
micro-controller maintains stability and eliminates noise pickup by
only outputting values to the DAC when the user changes the current
set point. The PIC18F4550 also includes a built-in analog to digital
converter (ADC) which we use to read back and display $V_{mon}$.

\subsection{Output Current Limitations}

To protect the laser diode, it is often desirable to limit the maximum
output current of the driver. An easy way to do this is by changing
the resistance of the sense resistor, $R_{s}$. Because the DAC can
only generate a set-point voltage as low as $V_{reg}-5V$, the maximum
output of the current driver will be $5\mbox{ V}/R_{s}$. However,
because the modulation current from the modulation input sub-circuit
does not pass through the sense resistor, it is possible to exceed
this maximum current by applying a large enough voltage to the modulation
input.

As discussed in \citep{Libbrecht93al}, another way to limit the output
current is by reducing the voltage $V_{reg}$. The compliance voltage
of the current driver is determined by the impedance of the mosfet
and the inductor in the current regulation sub-circuit and the value
of $V_{reg}$. If $I_{max}$ is the maximum current for the laser
diode, and if $V_{diode}$, $V_{mosfet}$, and $V_{L}$ are the voltage
drops across the laser diode, the mosfet, and the inductor (respectively)
at this current, then if $V_{reg}$ is set to\begin{equation}
V_{reg}=V_{diode}+V_{mosfet}+V_{L}+I_{max}R_{s},\label{eq:VregLimit}\end{equation}
the current driver output will not be able to exceed $I_{max}$. The
above voltage drops can be estimated from the specifications of the
components, or the ideal value for $V_{reg}$ can be determined empirically
by setting $V_{set}$ to its lowest value and gradually increasing
$V_{reg}$ until the output current equals $I_{max}$. Because an
inductor decouples the modulation input sub-circuit from the current
regulation sub-circuit, a high-frequency signal on the modulation
input could cause the current driver to output more than $I_{max}$.
At low modulation frequencies the output current will never exceed
$I_{max}$, regardless of the amplitude of the voltage applied to
the modulation input.

The maximum current that can be generated without risking damage to
the current driver is 127 mA (limited by the S1210 inductor). For
higher current applications, we designed our circuit boards to accommodate
both surface-mount and leaded components at a few critical points,
allowing a wide range of components to be used. For example, we have
realized a high current driver for a tapered amplifier by replacing
the following components: the S1210 inductor (rated for 127 mA) as
well as the two inductors in the \char`\"{}power filter\char`\"{}
sub-circuit were replaced with leaded 125 $\mu\mbox{H}$ inductors;
the NDS0605 mosfet (with a maximum rated current of 180 mA) was replaced
with an IRF9Z14 mosfet in a TO-220 package which can output 6.7 A;
the surface-mount LM317 regulator (with a current limit of 1 A) was
replaced with an LM1084-ADJ regulator in a TO-220 package, which is
capable of 5 A; The 1.5 Ohm resistor in the \char`\"{}voltage regulation\char`\"{}
sub-circuit was replaced with a leaded 1 $\Omega$, 5 W power resistor;
and the sense resistor $R_{s}$ was realized with two 5 $\Omega$,
10 W leaded power resistors placed in parallel for a combined resistance
of 2.5 $\Omega$. Due to the larger gate capacitance of the mosfet,
we also had to replace the capacitor in the \char`\"{}current regulation\char`\"{}
sub-circuit with a 100 nF capacitor to keep the circuit from oscillating.
Although these substitutions increase the noise and reduce the stability
of the circuit, they allow an output of 2 A (limited intentionally
by our choice of $R_{s}$ to match the capacity of our tapered amplifier).

\section{Results}

\label{sec:Results}We did a series of tests to characterize the performance
of our current driver. For each of these tests, the sense resistor
consisted of two 100 $\Omega$ resistors in parallel, for a sense
resistance of $R_{s}=50\,\Omega$.

\subsection{Noise and Stability}

\begin{figure}
\begin{centering}
\includegraphics{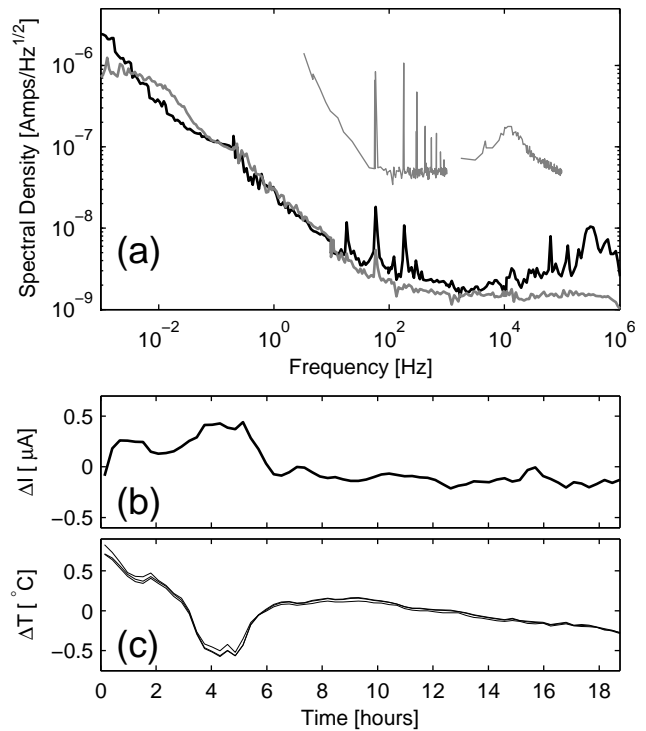} 
\end{centering}

\caption{Noise and drift in the current. In (a) the spectral density of the
current noise is shown. The black line represents the spectral density
of the current signal when the diode and resistor were driven by the
current driver. The thick gray line is the spectral density of the
intrinsic noise of the measurement apparatus, measured by disconnecting
the current driver from the laser diode. The thin gray lines are the
spectral density of the noise data in the original paper on the Hall-Libbrecht
design \citep{Libbrecht93al}. In (b) the measured drift of the output
current about its mean value is shown as a function of time. In (c)
the drift in the temperature at three locations is plotted (all three
sets of temperature data are plotted on top of each other). \label{cap:noise}}

\end{figure}

To characterize both the long-term stability and the high frequency
noise of the current driver, we used it to drive a laser diode in
series with a 2.5$\Omega$ precision resistor with a current of 74.5
mA. The current was determined by measuring the voltage drop across
the resistor with a 6.5 digit multimeter. From this data we extracted
the spectral density of the current noise, shown in figure \ref{cap:noise}(a).
The intrinsic noise of our measuring apparatus was determined by taking
a similar measurement but with the current driver disconnected from
the laser/2.5 $\Omega$ resistor. For most of the spectrum, our driver
is at or near the detection limit of our equipment, and at all frequencies
it is at least an order of magnitude smaller than the measurements
presented in \citep{Libbrecht93al}.

The long term drift in the output current is shown in Fig.\ \ref{cap:noise}(b).
This data was collected in the same manner as the data used to produce
Fig.\ \ref{cap:noise}(a), but with a current set point of 74.8 mA.
During this measurement thermistors were used to measure the temperature
at the location of the current sense resistor (labeled $Rs$ in Fig.\
\ref{cap:schematic}), the DAC, and the box that houses the current
driver. The mean temperatures at these locations was 40.6, 40.5, and
22.9$^{\circ}\mbox{C}$ respectively. As seen in Fig.\ \ref{cap:noise}(c),
the drift in current and temperature are highly correlated. This data
suggests a temperature coefficient of $\approx1.7\mbox{ ppm}/^{\circ}\mbox{C}$,
precisely what would be expected given the $2\mbox{ ppm}/^{\circ}\mbox{C}$
temperature coefficient of the sense resistor.

\subsection{Set-point Accuracy and Repeatability\label{sub:AccuracyAndRepeatability}}

\begin{figure}
\begin{centering}
\includegraphics{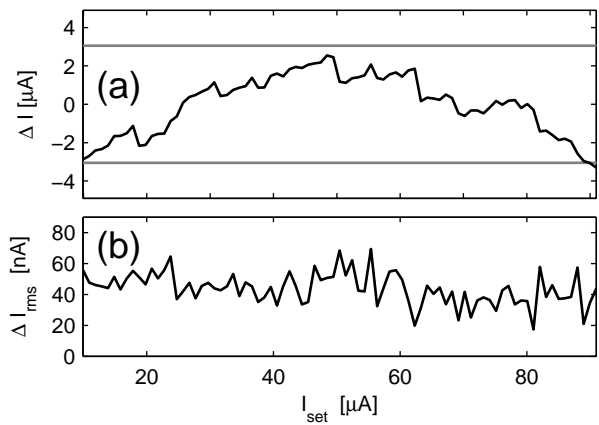} 
\end{centering}

\caption{Accuracy and repeatability of the current set point. The plot in (a)
shows the deviation of the actual current data from the set point.
The gray lines indicate the $\pm2$ LSB points specified in the DAC's
data sheet. Plot (b) show the rms repeatability of the current output,
measured as a function of the digitally programmed set point. \label{cap:accuracy}}

\end{figure}

The use of a digital set point results in very good accuracy, linearity,
and repeatability. We tested the linearity of the set point by connecting
our current driver to a laser diode in series with a 10 $\Omega$
precision resistor. We then determined the actual current produced
by the driver at various digital set points by measuring the voltage
drop across the resistor. The deviation of the actual current from
a best-fit line is shown in Fig.\ \ref{cap:accuracy}(a). The linearity
is within the $\pm2$ LSB specification for the DAC chip we used.
As discussed in Sec.\ \ref{sub:CurrentSetPoint}, similar chips with
higher or lower linearity are available. 

The set point repeatability of our circuit is much better than the
absolute accuracy. To test the repeatability, the current set point
was set digitally, and the actual current was measured. The digital
set point was then changed momentarily to significantly increase the
output current. The digital set point was then reset to its original
value, and the actual current was re-measured. This was done ten times
at each digital set point, and the rms deviation of those ten measurements
was calculated. The rms deviation is plotted as a function of the
digital current set point in Fig.\ \ref{cap:accuracy}(b). As shown
in this figure, the set-point repeatability is about 40 nA rms.

\subsection{Modulation Response}

\begin{figure}
\begin{centering}
\includegraphics{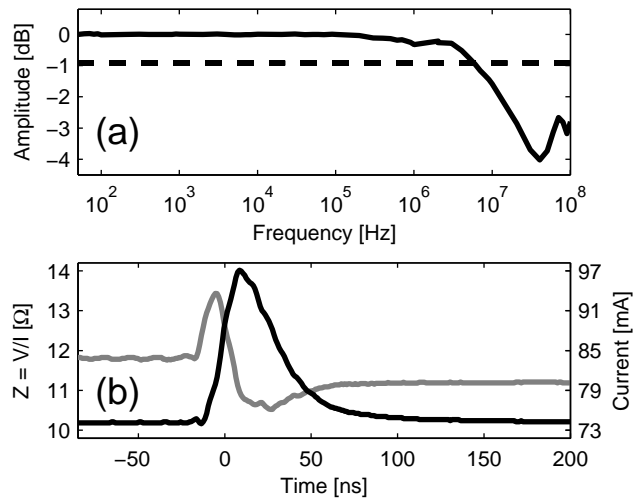} 
\end{centering}

\caption{Temporal response of the current driver. In (a) the Bode plot of the
response to the modulation input is displayed. The solid line is the
amplitude of the response of the driver normalized to the ideal $V_{mod}/R_{mod1}$
response for the signal applied to the modulation input. The dashed
line is the theoretical high frequency response assuming ideal components
and transmission lines given by Eq.\ \ref{eq:HFmodlimit}. In (b)
the transient response of the current driver to a total load change
of roughly 3\% is shown. The black line is the current through the
load, and the gray line is the impedance of the load. \label{cap:temporal}}

\end{figure}

We measured the modulation response of the current driver by terminating
the output with a 50 $\Omega$ resistor and comparing the amplitude
of a sine wave applied to the modulation input of the current driver
input to the amplitude of the sinusoidal voltage modulation generated
across this resistor. The output amplitude was then normalized to
the expected output of $V_{mod}/1\mbox{ k}\Omega$ to produce the
curve shown in Fig. \ref{cap:temporal}(a). We found the 3 dB point
to be over 20 MHz, with an attenuation of less than 4 dB out past
100 MHz. We found, however, that the frequency response was severely
degraded when driving reactive loads. As discussed in Sec.\ \ref{sub:CurrentModulation},
this is a much better frequency response than one might naively expect
based on the bandwidth of the op-amps.

\subsection{Transient Response to Load Impedance Variation}

The transient response of the current driver was studied by using
it to drive current through a 10 $\Omega$ resistor in series with
a VZN1206 mosfet. The transistor gate voltage was suddenly changed
to vary the overall impedance of the load seen by the current driver.
The current was determined from the voltage across the resistor, and
the time-varying impedance was calculated by dividing the voltage
drop across the entire transistor-resistor circuit by the current
found in the resistor. The measurements are shown in Fig.\ \ref{cap:temporal}(b).
The distance between the peaks in the two traces is 14 ns, indicating
a very fast initial response to the impedance change. After 100 ns,
the impedance had settled down to a constant value. By fitting the
current signal beyond this time to an exponential, we extracted a
time constant for the driver's response to small load changes of 70
ns.

\section{Conclusion}

Our laser diode current driver design, an improvement upon the Hall-Libbrecht
design \citep{Libbrecht93al}, has extremely good long-term stability,
low noise, fast modulation response, and fast response to load impedance
changes. The implementation of a digital set-point circuit and the
use of modern surface-mount chips has resulted in more than an order-of-magnitude
improvement in noise and stability. It also allows for extremely precise
repeatability, high accuracy, and the ability to control the current
set point remotely without degrading the performance of the current
driver. The device out-performs any commercial device we are aware
of, yet is simple and inexpensive enough to be used as a {}``lab
standard'' for both highly sensitive as well as less sensitive diode
laser applications.

This work was supported by a grant from the Research Corporation and
by Brigham Young University's Office of Research and Creative Activities.

\end{document}